\documentclass[conference]{IEEEtran}
\ifCLASSINFOpdf
\else
\fi
\usepackage{amsmath}
\usepackage{amssymb}

\usepackage{graphicx}

\begin{document}
\title{{\huge Optimal Binary Coding for $q^+$-state Data Embedding}}
\author{\IEEEauthorblockN{Han-Zhou Wu}
\IEEEauthorblockA{E-mail: wuhanzhou\_2007@126.com}}


\maketitle


\begin{abstract}
In steganography, we always hope to maximize the embedding payload subject to an upper-bounded distortion. 
We need suitable distortion measurement to evaluate the embedding impact. 
However, different distortion functions exposes different levels of distortion evaluation, implying that
we have different optimization distributions by applying different distortion functions. 
In applications, the embedding distortion is caused by a certain number of embedding operations. 
Instead of a predefined distortion, we actually utilize a number of modifications 
to embed as many message bits as possible as long as the modifications are acceptable. 
This paper focuses on the design of optimal binary codewords for data embedding with a limited number of modification candidates. 
We have proved the optimality of the designed codewords, and proposed the way to construct the optimal binary codewords. 
It is pointed out that the optimal binary code is not unique, and an optimal code can be computed within a low computational cost.
\end{abstract}


%
\IEEEpeerreviewmaketitle

\section{Motivation}
Steganography [1] refers to embed secret bits into innocent signals, e.g., digital images, 
by slightly altering the insignificant components of cover signals for covert communication. 
It is desirable to hide as many secret bits as possible without introducing statistically detectable artifacts into a cover object. 
This can be generally modeled by the payload-distortion performance. One expects to minimize the heuristically-defined distortion 
for a \emph{lower-bounded payload} (LBP), while one may require to maximize the embedding payload subjected to an \emph{upper-bounded distortion} (UBD).

For the LBP problem, a way to evaluate embedding (coding) algorithm in steganography is to compare the embedding efficiency (in bits per unit distortion) for a fixed expected relative payload. 
A higher embedding efficiency often implies a lower distortion can be achieved. To maximize the embedding payload (i.e., the UBD problem), an effective way to evaluate embedding algorithm is to analyze the embedding redundancy, 
which reveals the modification utilization during data embedding process. In applications, a lower embedding redundancy is required 
for an upper-bounded distortion. Though the data embedding operation are modeled as two forms (i.e., the LBP and UBD), 
they are dual to each other, which indicates that the optimal statistical distribution for the LBP may be optimal to the UBD as well, for some upper-bounded distortion [2].

Many practical algorithms based on coding theory have been proposed since the embedding efficiency of steganographic schemes (corresponds to the LBP problem) 
can be improved by applying covering codes. A well-known technique is matrix embedding [3], where the sender minimizes the total number of embedding changes for a fixed relative payload, 
resulting in a high embedding efficiency. Other relative-optimal covering codes such as hamming codes [4], BCH codes [5], wet paper codes [6] and syndrome-trellis codes [2], are introduced in the literature. 
These novel covering codes mainly focus on embedding while minimizing the average distortion subjected to a fixed relative payload, which provide an effective way for a payload-limited steganographic system.

On the other hand, the UBD problem corresponds to a more intuitive use of steganography since cover signals with a different level of noise or texture can carry a different level of embedding payload. 
It indicates that the distortion should be constrained instead of payload, to maximize the amount of hidden information. 
For a specified cover object, one may expect to embed as many message bits as possible as long as the distortion corresponds to an acceptable statistical detectability. 
Though the determined function between the distortion and the statistical detectability is unknown currently, data embedding while minimizing the distortion function is desirable. 
It means that we expect to maximize the expected relative payload for a specified cover object constrained by a flexible upper-bounded average distortion. 
Thereafter, we may further expect to reduce the upper-bounded average distortion under the condition that the expected relative payload can be achieved. 
Based on this perspective, in this paper, instead of modeling the embedding distortion by some specified function, we consider the number of embedding modifications. 
We aim to present a coding technique for embedding as high payload as possible with the limited number of distorted modifications.

\section{Preliminary Concepts}
\subsection{Problem Formulation}
Without the loss of generality, we will call a sequence of $n$ elements $\mathbf{x}=(x_1,x_2,...,x_n)\in X=X_1\times X_2\times ...\times X_n=\{0,1,...,2^d-1\}^n$ the cover object,
where $d$ is the number of bits needed to describe each element. For example, we can consider the cover object as a digital image, e.g., $d$ = 8 for an 8-bit grayscale image. 
The data sender communicates a message $\mathbf{m}\in M$, where $M$ is the set of all messages $\mathbf{m}$ that can be communicated, to the data recipient 
by introducing modifications to the cover image and sending the corresponding stego image $\mathbf{y}=(y_1,y_2,...,y_n)\in Y=Y_1\times Y_2\times ...\times Y_n=\{0,1,...,2^d-1\}^n$.
For steganography, it requires
\begin{equation}
\forall \mathbf{x}\in X, \mathbf{m}\in M, \exists \mathbf{y}\in Y,Ext(\mathbf{y})=Ext(Emb(\mathbf{x},\mathbf{m}))=\mathbf{m}.
\end{equation}
where both the data embedding and data extraction may utilize a secret key to improve the steganographic security.

It is true that different embedding rates result in different distortion between the cover image and stego one. The impact of making embedding changes at cover elements can be measured by using some 
heuristically-defined distortion function $D(\mathbf{x},\mathbf{y})=||\mathbf{x}-\mathbf{y}||_D$, e.g., measuring an embedding change using a cost scalar.
To protect the secret message securely, we assume the data sender obtains the embedding payload in the form of a pseudo-random message bit stream, such as by encrypting the original message with a cryptographic method. 
It means that the message $\mathbf{m}\in M$ is always considered as a pseudo-random bit stream. 

The data embedding algorithm associates a specified cover image $\mathbf{x}$ with a pair $\{Y, \pi\}$ where $Y$ is the set of all stego images 
into which $\mathbf{x}$ can be modified and $\pi$ describes their probability distribution satisfying $\pi(\mathbf{y}) = \mathrm{Pr}\{Y = \mathbf{y} | \mathbf{x}\}$. 
For simplicity, we here consider $\mathbf{x}$ as a constant parameter that is fixed in the very beginning and we do not further denote the dependency on it explicitly. 
Therefore, we simply replace the embedding distortion $D(\mathbf{x},\mathbf{y})$ with $D(\mathbf{y})$, namely $D(\mathbf{y}) = D(\mathbf{x},\mathbf{y})$. 
If the data recipient knew $\mathbf{x}$, the data sender could send up
\begin{equation}
H(\pi)=\sum_{\mathbf{y}\in Y}\pi(\mathbf{y})\cdot \mathrm{log}_2\frac{1}{\pi(\mathbf{y})}.
\end{equation}
bits on average to the data recipient while introducing the average distortion
\begin{equation}
E(D_{\pi})=\sum_{\mathbf{y}\in Y}\pi(\mathbf{y})\cdot D(\mathbf{y}).
\end{equation}
by selecting the stego image according to $\pi$. In practice, we are interested in practical methods that can embed at least $m$-bit message in an $n$-element cover object, 
while keeping the expected distortion as small as possible. We think of it as the LBP problem, which specifies the optimization problem
\begin{equation}
\underset{\pi}{\mathrm{arg~min}}~E(D_{\pi}),~\mathrm{subject~to}~H(\pi)\geq m.
\end{equation}

On the other hand, one expect to embed as many message bits as possible while introducing a limited average distortion. 
We think of this as the UBD problem, which specifies the optimization problem
\begin{equation}
\underset{\pi}{\mathrm{arg~max}}~H(\pi),~\mathrm{subject~to}~E(D_{\pi})\leq \rho.
\end{equation}

For the LBP problem, one has to define a distortion function for significantly describing distortion characteristics due to data embedding. 
It implies that we may have different optimal distributions when to utilize different distortion functions. 
Compared with the LBP problem, the UBD problem corresponds to a more intuitive use of steganography. 
We will focus on the optimization of the UBD problem. 
Though the UBD problem considers the constraint of the embedding distortion, the hidden information are actually 
carried by the cover pixels according to a number of modification candidates, meaning that, 
instead of considering the constraint of a heuristically-defined distortion function, 
we are to maximize the embedding payload based on a number of pixel modifications 
as long as the pixel modifications are usable (namely, the modification candidates are acceptable). 
We will introduce this viewpoint in detail in the following subsection.
\subsection{Entropy Bound and Redundancy Metric}
In steganography, cover elements are generally divided into disjoint blocks to respectively carry additional information. 
For consistency, we replace the cover vector $\mathbf{x}=(x_1,x_2,...,x_n)$ with $\mathbf{x}=(\mathbf{x}^{(1)},\mathbf{x}^{(2)},...,\mathbf{x}^{(s)})$, where
$s$ denotes the number of blocks and $\mathbf{x}^{(k)}=(x_{(k-1)\cdot r+1}, x_{(k-1)\cdot r+2}, ..., x_{kr}),~(1\leq k\leq s, n=r\cdot s)$.
During data embedding, the data sender selects the corresponding stego vector $\mathbf{y}=(\mathbf{y}^{(1)}, \mathbf{y}^{(2)}, ..., \mathbf{y}^{(s)})$ to 
carry the secret information $\mathbf{m}=(\mathbf{m}^{(1)}, \mathbf{m}^{(2)}, ..., \mathbf{m}^{(s)})$. Here, we say $\mathbf{y}^{(k)}~(1\leq k\leq s)$ has 
the identical number of elements with $\mathbf{x}^{(k)}$ and each $\mathbf{m}^{(k)}~(1\leq k\leq s)$ corresponds to a random bit stream with an indefinite length. 
It can be seen from Eq. (5) that the optimization is to find such a stego vector distribution that the amount of payload is the highest 
while the expected embedding impact should be no more than a threshold. 
To evaluate the embedding impact, we generally use some suitable distortion measurement such as mean absolute error (MAE) and mean square error (MSE). 
However, a different distortion function results in a different level of distortion evaluation. 
It implies we may have different optimization distributions by applying different distortion functions. 
In practical applications, the embedding distortion is caused by a number of embedding modifications. 
Therefore, for the UBD problem, instead of considering the constraint of a predefined distortion, 
we actually can utilize a number of modifications to embed as many message bits as possible 
as long as all the modification states correspond to a tolerable distortion. 

Specifically, for each block $\mathbf{x}^{(k)}~(1\leq k \leq s)$, according to the data embedding operation, 
we know the amount of all possible $\mathbf{y}^{(k)}$, denoted by $|S(\mathbf{y}^{(k)})|$, 
where $S(\mathbf{y}^{(k)})$ represents the block set containing all possible $\mathbf{y}^{(k)}$ derived from $\mathbf{x}^{(k)}$. 
To approach the UBD problem, we expect to maximize the expected embedding payload for each block $\mathbf{x}^{(k)}~(1\leq k \leq s)$ by modifying $\mathbf{x}^{(k)}$ 
as one of the $|S(\mathbf{y}^{(k)})|$ resultant states in practical applications. 
It can be seen that the goal is to maximize the expected bit length of $\mathbf{m}^{(k)}~(1\leq k \leq s)$, denoted by $l(\mathbf{m}^{(k)})$. 
Therefore, the UBD problem can be described as another form
\begin{equation}
\underset{\mathbf{x}~\mapsto~\mathbf{y}}{\mathrm{arg~max}}\sum_{k=1}^{s}l(\mathbf{m}^{(k)}),~\mathrm{subject~to}~|S(\mathbf{y}^{(k))})|=q_k~(1\leq k\leq s).
\end{equation}
where $q_1, q_2, ..., q_s$ denote the amount of all possible stego states. Since, in practical applications, 
the data embedding process for any two cover blocks $\mathbf{x}^{(i)}$ and $\mathbf{x}^{(j)}$ are generally independent of each other and 
utilize the identical data embedding function, we think of $l(\mathbf{m}^{(1)})$, $l(\mathbf{m}^{(2)})$, ..., $l(\mathbf{m}^{(s)})$ as identical, 
and $q_1$, $q_2$, ..., $q_s$ as well. For simplicity, we assume $l(\mathbf{m}^{(1)})=l(\mathbf{m}^{(2)})=...=l(\mathbf{m}^{(s)})=l^+$ and $q_1=q_2=...=q_s=q^+$. 
Based on the entropy theory, the expected embedding payload satisfies
\begin{equation}
H(\pi)=\sum_{k=1}^{s}l(\mathbf{m}^{(k)})=\sum_{k=1}^{s}l^+=\frac{n}{r}\cdot l^+\leq \frac{n\cdot \mathrm{log}_2q^+}{r}.
\end{equation}

In order to embed as many message bits as possible, we expect to find a coding algorithm such that the embedding payload 
nears to the theoretical bound as shown in Eq. (7). In applications, since both $n$ and $r$ can be pre-determined, 
we are to find a coding algorithm such that each cover block can carry a payload that nears to the payload bound, i.e., $\mathrm{log}_2q^+$. 
Note that, \emph{any steganographic scheme can be considered as a special case meeting that} $n = r,~s=1$.
An effective way of evaluating coding algorithms is to compare the embedding redundancy. 
Thus, based on the Eq. (7), the embedding redundancy $\eta$ here is formulated as
\begin{equation}
\eta=1-\frac{H(\pi)}{H_{\textrm{max}}(\pi)}=1-\frac{\sum_{k=1}^{s}l(\mathbf{m}^{(k)})}{n/r\cdot \mathrm{log}_2q^+}=1-\frac{l^+}{\mathrm{log}_2q^+}.
\end{equation}

Generally, a lower embedding redundancy implies a better modification utilization, 
which results in a higher embedding payload for the coding algorithm. 
It can be seen from Eq. (8) that we need to design a coding algorithm for each cover block such that $l^+$ 
is maximal in order to minimize the embedding redundancy for a fixed $q^+$. It should be noted that, we here assume that, the data embedding operations to any 
two cover blocks are independent of each other. In the following section, we are to introduce such a coding algorithm called \emph{optimal binary coding (OBC)} 
that minimizes the redundancy.

\section{Optimal Binary Coding}
We introduce a coding technique for minimizing the embedding redundancy in this section. 
For each $\mathbf{x}^{(k)}~(1\leq k \leq s)$, the amount of all resultant possible $\mathbf{y}^{(k)}$ is 
$|S(\mathbf{y}^{(k)})|$, where $S(\mathbf{y}^{(k)})$ = $\{{\mathbf{y}_1}^{(k)}, {\mathbf{y}_2}^{(k)}, ..., {\mathbf{y}_{q^+}}^{(k)}\}$ represents 
the block set containing all possible $\mathbf{y}^{(k)}$ derived from $\mathbf{x}^{(k)}$.
During data embedding, $\mathbf{x}^{(k)}$ will be replaced with an element in $S(\mathbf{y}^{(k)})$ to carry a prefix of the secret data. 
Note that $|S(\mathbf{y}^{(k)})|=q^+$ and $q^+\geq 2$. 
We expect to find an optimal mapping function $F:~\mathbf{y}^{(k)}\rightarrow\mathbf{m}^{(k)}$. 
It is required that, for an arbitrary bit stream $\mathbf{x}^{(k)}~(1\leq k \leq s)$, there exists at least one 
$\mathbf{m}^{(k)} \in \{{\mathbf{m}_1}^{(k)}, {\mathbf{m}_2}^{(k)}, ..., {\mathbf{m}_{q^+}}^{(k)}\}$ that 
is a prefix of the secret data since the secret data can be any bit stream, i.e.,
\begin{equation}
\forall \mathbf{m}\in M,\exists i\in [1,q^+], {\mathbf{m}_i}^{(k)}\in \mathrm{Pre}(\mathbf{m}).
\end{equation}
where Pre($\mathbf{m}$) denotes the set that contains all the prefixes of $\mathbf{m}$, e.g., 
Pre(``0110'') = \{``0'', ``01'', ``011'', ``0110''\}. It indicates that, the cover block should always 
be altered to match a prefix of the bit stream to be embedded.

Without the loss of generality, we think of ${\mathbf{m}_i}^{(k)}$ as the assigned bit stream for 
${\mathbf{y}_i}^{(k)}~(1\leq i\leq q^+)$, namely ${\mathbf{m}_i}^{(k)}=F({\mathbf{y}_i}^{(k)})$, $1\leq i\leq q^+$. 
Let $l({\mathbf{m}_i}^{(k)})~(1\leq i\leq q^+)$ denote the bit length of the assigned bit stream ${\mathbf{m}_i}^{(k)}$. 
It can be seen that
\begin{equation}
l^+=\sum_{i=1}^{q^+}\mathrm{Pr}\{{\mathbf{y}}^{(k)}={\mathbf{y}_i}^{(k)}|{\mathbf{x}}^{(k)}\}\cdot l({\mathbf{m}_i}^{(k)}).
\end{equation}
where
\begin{equation}
\sum_{i=1}^{q^+}\mathrm{Pr}\{{\mathbf{y}}^{(k)}={\mathbf{y}_i}^{(k)}|{\mathbf{x}}^{(k)}\}=1.
\end{equation}

As shown in Eq. (8), in order to minimize the embedding redundancy, we expect to obtain the maximum $l^+$ when the amount of stego states, i.e. $q^+$, is fixed. 
In the following, we introduce a technique to find such a bit stream mapping function that ensures a maximum $l^+$ for a fixed $q^+$.

For a fixed $q^+$, to make full use of all the $q^+$ stego states, a basic restriction is imposed on a bit stream mapping function for steganography: 
\emph{no two bit streams satisfy that one is a prefix of the other}. It means that, for $F:~\mathbf{y}^{(k)}\rightarrow\mathbf{m}^{(k)}$, 
there does not exist such an index-pair $(i, j)~(1\leq i \neq j\leq q^+)$ such that ${\mathbf{m}_i}^{(k)}$ is a prefix of ${\mathbf{m}_j}^{(k)}$. 
Since Eq. (9) should hold for the mapping function \emph{F}, 
assuming that there exists an index-pair $(i, j)~(1\leq i \neq j\leq q^+)$ such that ${\mathbf{m}_i}^{(k)}$ is a prefix of ${\mathbf{m}_j}^{(k)}$, 
it can be seen that when the secret message starts from ${\mathbf{m}_j}^{(k)}$, it also starts from ${\mathbf{m}_i}^{(k)}$, 
which means that ${\mathbf{m}_j}^{(k)}$ can be replaced by ${\mathbf{m}_i}^{(k)}$ resulting in that ${\mathbf{m}_j}^{(k)}$ will never be used. 
Thus, there will actually be $(q^+-1)$ stego states used for steganography since ${\mathbf{y}_j}^{(k)}$ can be always replaced by ${\mathbf{y}_i}^{(k)}$ 
to match a prefix of the secret message, which implies that the mapping function does not make full use of all the $q^+$ stego states. 
It can be inferred from the basic restriction that no two stego states map to an identical bit stream since a bit stream must be a prefix of itself. 
\begin{figure}[!t]
\centering
\includegraphics[width=3.0in]{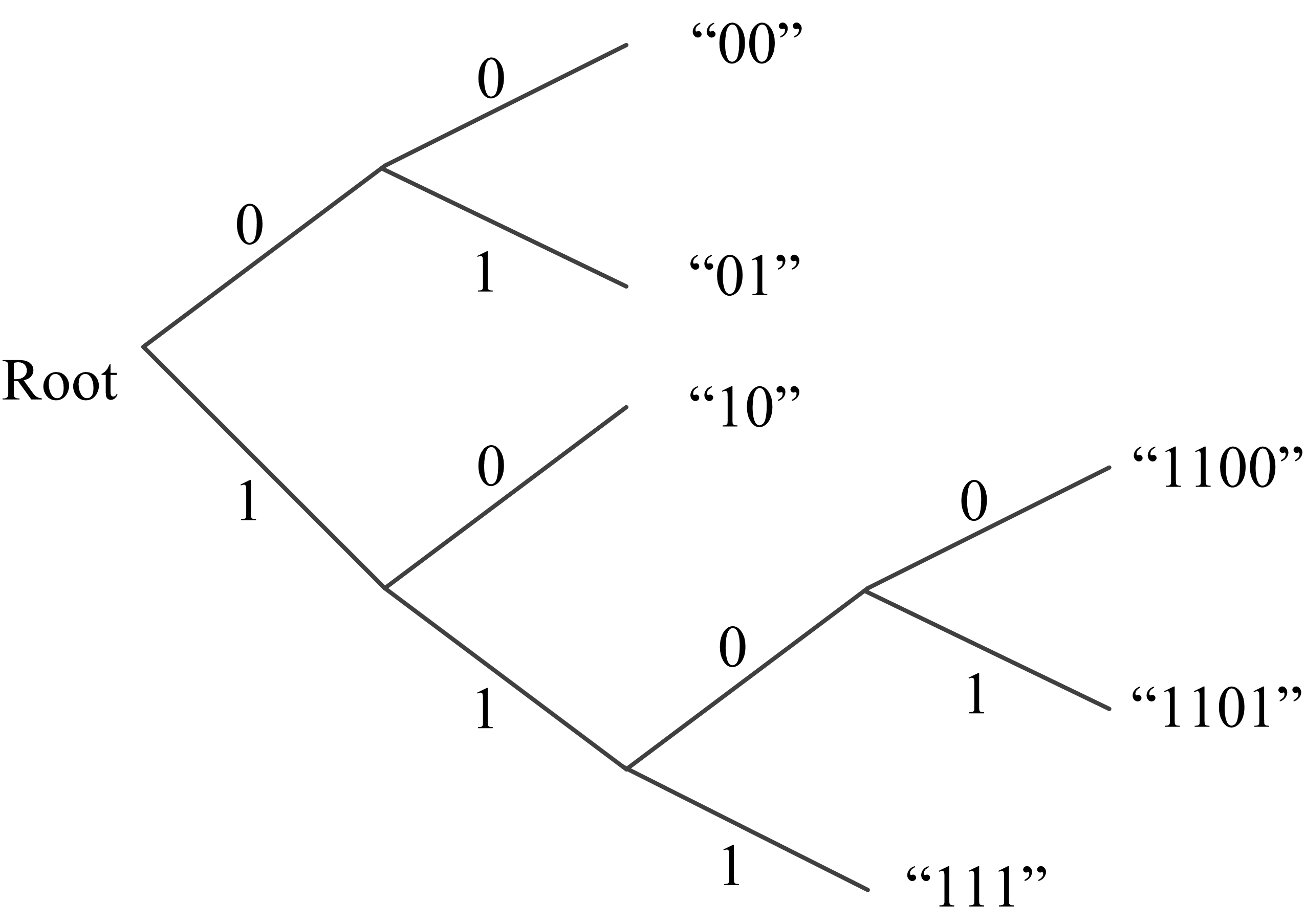}
\caption{An example for the binary prefix code.}
\end{figure}

The bit stream mapping function \emph{F} is equivalent to a coding approach. Only one codeword in $\{{\mathbf{m}_1}^{(k)}, {\mathbf{m}_2}^{(k)}, ..., {\mathbf{m}_{q^+}}^{(k)}\}$
will match a prefix of the secret data. We need to construct an \emph{instantaneous code} (also named as a \emph{prefix code}) [7] to 
ensure the data embedding process. As $\{{\mathbf{m}_1}^{(k)}, {\mathbf{m}_2}^{(k)}, ..., {\mathbf{m}_{q^+}}^{(k)}\}$ constitute an instantaneous code, 
the probability of utilizing ${\mathbf{y}_i}^{(k)}$ to carry additional information equals the probability of that ${\mathbf{m}_i}^{(k)}$ matches 
a prefix of the secret data to be embedded. Since the secret data to be embedded forms a pseudo-random bit stream, 
it means the probability of that ${\mathbf{m}_i}^{(k)}$ matches a prefix of the secret data equals the probability of that 
${\mathbf{m}_i}^{(k)}$ matches a prefix of a pseudo-random bit stream. 
The probability of that ${\mathbf{m}_i}^{(k)}$ matches a prefix of a pseudo-random bit stream is $1/ 2^{l({\mathbf{m}_i}^{(k)})}$. 
Therefore
\begin{equation}
\mathrm{Pr}\{{\mathbf{y}}^{(k)}={\mathbf{y}_i}^{(k)}|{\mathbf{x}}^{(k)}\}=2^{-l({\mathbf{m}_i}^{(k)})},~(1\leq i\leq q^+).
\end{equation}
Then, Eq. (10) can be derived as
\begin{equation}
l^+=\sum_{i=1}^{q^+}l({\mathbf{m}_i}^{(k)}) \cdot 2^{-l({\mathbf{m}_i}^{(k)})}.
\end{equation}

We wish to construct an instantaneous code with the maximum expected length. 
This is equivalent to finding such a code $C$ that the expected length $l^+$ is maximum, which corresponds to a standard optimization problem
\begin{equation}
\underset{C}{\mathrm{arg~max}}~l^+,~\mathrm{subject~to}~\sum_{i=1}^{q^+}2^{-l({\mathbf{m}_i}^{(k)})}=1.
\end{equation}

An important property of an optimal code is determined to construct the optimal code. 
It specifies that the difference between the bit length of the longest codeword and that of the shortest codeword should be no more than one, 
i.e., for an optimal code $C=\{{\mathbf{m}_1}^{(k)}, {\mathbf{m}_2}^{(k)}, ..., {\mathbf{m}_{q^+}}^{(k)}\}$, it satisfies
\begin{equation}
\mathrm{max}\{l({\mathbf{m}_i}^{(k)})-l({\mathbf{m}_j}^{(k)}),~1\leq i,j\leq q^+\}\leq 1.
\end{equation}
\textbf{Proof.}

Assume that we have found such a prefix code $A = \{a_1, a_2, ..., a_{q^+}\}~(q^+ \geq 2)$ that the difference between the bit length of the longest 
codeword $a_j$ and that of the shortest codeword $a_i$ is higher than one, denoted as $l_j - l_i \geq 2$. 
We denote $a_j$ and $a_i$ as the form of a bit stream ``$b_1b_2...b_{l_j-1}b_{l_j}$'' and ``$c_1c_2...c_{l_i-1}c_{l_i}$'', respectively. 
Since the codeword $a_j$ has a longest bit length, for ensuring that $A$ is a prefix code, there must exist such an index $1 \leq k \leq q^+$ 
that $a_k$ = ``$b_1b_2...b_{l_j-1}(1-b_{l_j})$'' $\in A$. We compute the expected bit length of the prefix code $A$ as
\begin{equation}
{l^+}_A=\sum_{t\neq i,j,k}(l_t\cdot 2^{-l_t})+l_i\cdot 2^{-l_i}+l_j\cdot 2^{-l_j+1}.
\end{equation}

It can be inferred that both the bit stream ``$c_1c_2...c_{l_i-1}c_{l_i}0$'' and ``$c_1c_2...c_{l_i-1}c_{l_i}1$'' are not a codeword of $A$ since 
the codeword $a_i$ = ``$c_1c_2...c_{l_i-1}c_{l_i}$'' matches a prefix of both the bit stream ``$c_1c_2...c_{l_i-1}c_{l_i}0$'' and 
``$c_1c_2...c_{l_i-1}c_{l_i}1$''. Similarly, the bit stream ``$b_1b_2...b_{l_j-1}$'' is not a codeword of $A$ 
since ``$b_1b_2...b_{l_j-1}$'' is a prefix of $a_j$. This way, we can construct a new prefix code $A^\circ$ by replacing the three codewords 
$a_i$, $a_j$, and $a_k$ with ``$c_1c_2...c_{l_i-1}c_{l_i }0$'', ``$c_1c_2...c_{l_i-1}c_{l_i}1$'', and $b_1b_2...b_{l_j-1}$'', respectively. 
Therefore, we have
\begin{equation}
{l^+}_{A^\circ}=\sum_{t\neq i,j,k}(l_t\cdot 2^{-l_t})+(l_i+1)\cdot 2^{-(l_i+1)+1}+(l_j-1)\cdot 2^{-(l_j-1)}.
\end{equation}
Thus,
\begin{equation}
\Delta ={l^+}_{A^\circ}-{l^+}_{A}=\frac{1}{2^{l_i}}-\frac{1}{2^{l_j-1}}.
\end{equation}

As $l_j-l_i\geq 2$, it can be seen that $\Delta > 0$, which means that we can always construct a prefix code $A^\circ$ with 
a larger expected bit length by modifying the prefix code $A$ that does not meet Eq. (15). Therefore, it can be inferred that Eq. (15) holds for 
the optimal prefix code (instantaneous code). $~\square$

Generally, a binary prefix code corresponds to a binary tree in which each node has two children. 
Let the edges of the tree represent the symbols (``0'' and ``1'') for the prefix code. For example, the two edges arising from the root node 
represent the two possible values of the first symbol (``0'' or ``1'') for the prefix code. 
Each codeword is represented by a leaf on the tree. Fig. 1 shows an example for the binary prefix code. 
The prefix condition on the codewords implies that no codeword is an ancestor of any other codeword on the tree. 
Hence, each codeword eliminates its descendants as possible codewords.

The height of a node is the number of edges from the root to the node. 
Thus, the root has a height of zero. As each node on a binary tree has at most two branches, 
the amount of nodes with a height of $h$ is at most $2^h$. 
It can be seen that the bit length of a codeword is equal to the height of the corresponding node on the binary tree. 
Since Eq. (15) holds for the optimal prefix code $C$ with $q^+$ codewords, we have
\begin{equation}
\mathrm{min}\{l({\mathbf{m}_i}^{(k)}),~1\leq i\leq q^+\}=\left \lfloor \mathrm{log}_2q^+ \right \rfloor.
\end{equation}

For an optimal code $C$, let $n_1$ and $n_2$ denote the amount of codewords corresponding to a node with a height of $\left \lfloor \mathrm{log}_2q^+ \right \rfloor$ and 
the amount of codewords corresponding to a node with a height of ($\left \lfloor \mathrm{log}_2q^+ \right \rfloor$+1), respectively. We have
\begin{equation}
n_1+n_2=q^+.
\end{equation}

For the optimal code $C$, the amount of tree-nodes with a height of $\left \lfloor \mathrm{log}_2q^+ \right \rfloor$ is 
$2^{\left \lfloor \mathrm{log}_2q^+ \right \rfloor}$ , we further have
\begin{equation}
n_1+\frac{n_2}{2}=2^{\left \lfloor \mathrm{log}_2q^+ \right \rfloor}.
\end{equation}
namely,
\begin{equation}
n_1=2^{\left \lfloor \mathrm{log}_2q^+ \right \rfloor+1}-q^+,n_2=2q^+-2^{\left \lfloor \mathrm{log}_2q^+ \right \rfloor+1}.
\end{equation}

Therefore, both the amount of codewords with a bit length of $\left \lfloor \mathrm{log}_2q^+ \right \rfloor$ and the amount of codewords with a bit length of ($\left \lfloor \mathrm{log}_2q^+ \right \rfloor$+1) are 
uniquely determined for the fixed $q^+$. It can be further determined form Eq. (13) that the expected bit length of the optimal code is 
also uniquely determined that
\begin{equation}
{l^+}_{\mathrm{max}}=\left \lfloor \mathrm{log}_2q^+ \right \rfloor+\frac{q^+}{2^{\left \lfloor \mathrm{log}_2q^+ \right \rfloor}}-1.
\end{equation}

Therefore, the optimal prefix code $C$ with $q^+$ codewords corresponds to the minimum embedding redundancy $\eta_{\mathrm{min}}$:
\begin{equation}
\eta_{\mathrm{min}}=1-\frac{{l^+}_{\mathrm{max}} }{{\mathrm{log}_2}q^+}
=\frac{({\mathrm{log}_2}q^+-\left \lfloor \mathrm{log}_2q^+ \right \rfloor+1)\cdot 2^{\left \lfloor \mathrm{log}_2q^+ \right \rfloor}-q^+}{{\mathrm{log}_2}q^+\cdot 2^{\left \lfloor \mathrm{log}_2q^+ \right \rfloor}}.
\end{equation}

Fig. 2 shows the comparison between the theoretical bound and the proposed OBC in terms of the embedding capacity.
It is observed that, the curve of OBC is rather close to that of the theoretical bound. Based on the above analysis, 
we are to construct the codewords for $C=\{{\mathbf{m}_1}^{(k)}, {\mathbf{m}_2}^{(k)}, ..., {\mathbf{m}_{q^+}}^{(k)}\}$. 
At first, we collect all binary codewords with a length of exactly $\left \lfloor \mathrm{log}_2q^+ \right \rfloor$, denoted by $C_q$, e.g., 
if $q^+$ = 6, the collected codewords will be $C_q$ = \{``00'', ``01'', ``10'', ``11''\}. 
Then, according to Eq. (22), we randomly select $n_2/2$, i.e., $q^+-2^{\left \lfloor \mathrm{log}_2q^+ \right \rfloor}$, codewords from 
$C_q$ to generate $n_2$ new codewords by appending ``0'' and ``1'' to the end of the selected codewords. 
And, the rest non-selected codewords are kept unchanged to constitute $n_1$ new codewords. In this way, we can finally construct the codewords for $C$.
For example, if $q^+$ = 6, we can select \{``00'', ``01''\} out from $C_q$, and obtain \{``000'', ``001'', ``010'', ``011''\} by appending ``0'' and 
``1'' to the end of ``00'' and ``11''. We then use \{``10'', ``11''\} in $C_q$ and \{``000'', ``001'', ``010'', ``011''\} to 
construct the codewords as $C$ = \{``000'', ``001'', ``010'', ``011'', ``10'', ``11''\}. Therefore, we can finally construct the required OBC codewords.
\begin{figure}[!t]
\centering
\includegraphics[width=3.0in]{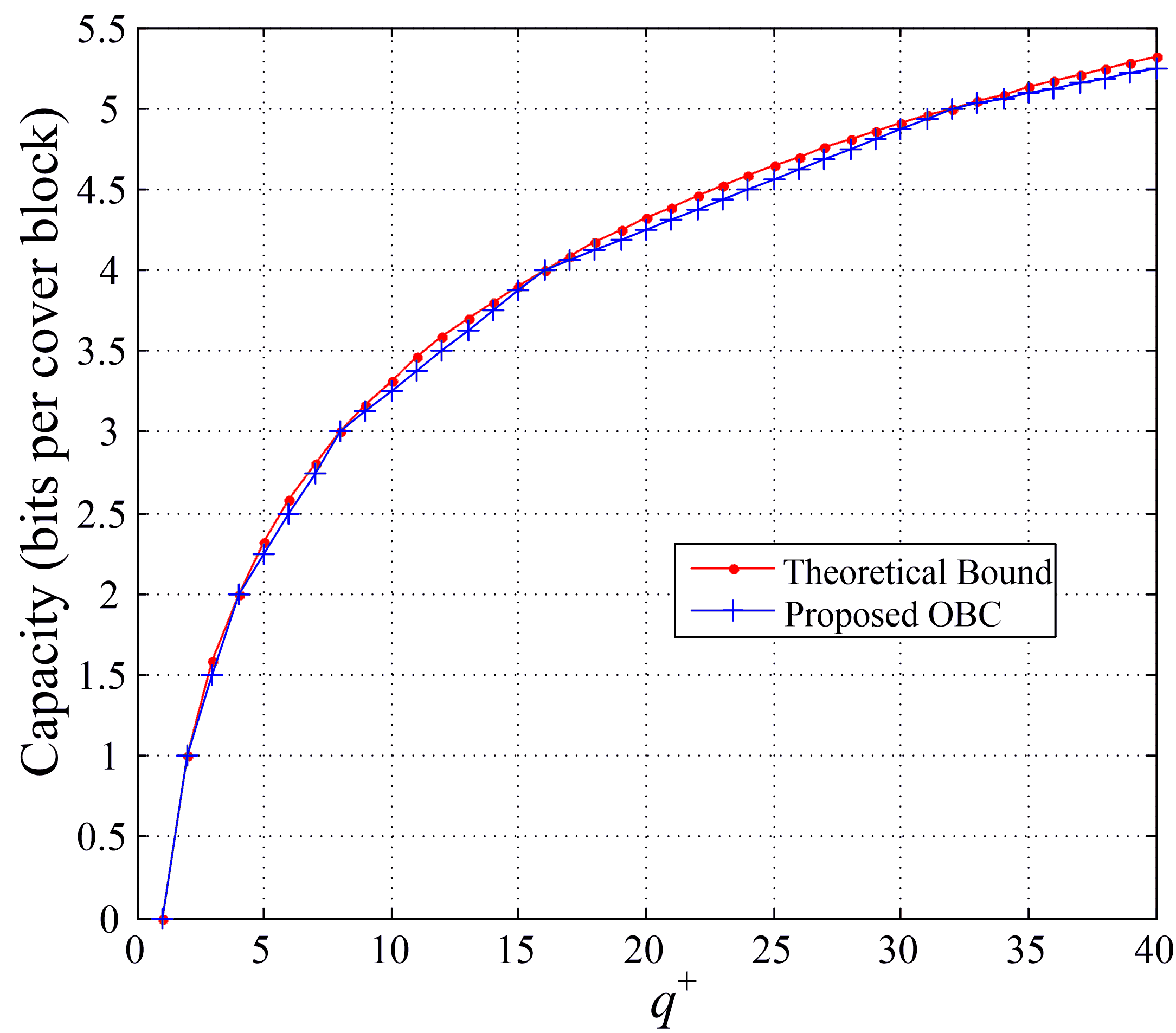}
\caption{Comparison in terms of the embedding capacity.}
\end{figure}

\section{Discussion}
In this paper, we present the OBC technique by modeling data embedding as optimal coding problem for a limited number of modification candidates.
It is noted that, the OBC code is not unique since we can append ``0'' and ``1'' to the end of arbitrary $q^+-2^{\left \lfloor \mathrm{log}_2q^+ \right \rfloor}$ 
codewords in $C_q$. In applications, to quickly determine each ${\mathbf{m}_i}^{(k)},~(1\leq i\leq q^+)$, we can apply the appending operation to the 
smallest $n_2/2$ codewords of $C_q$ in the form of decimal notation. Thus, the value of ${\mathbf{m}_i}^{(k)}$ can be computed within a computational 
cost of $O(\mathrm{log}_2q^+)$. On the other hand, we may hope to minimize $E(D({\mathbf{y}_i}^{(k)}))$ when to apply the OBC technique. 
It requires us to rearrange the index-mapping between $\{{\mathbf{y}_1}^{(k)}, {\mathbf{y}_2}^{(k)}, ..., {\mathbf{y}_{q^+}}^{(k)}\}$ and 
$\{{\mathbf{m}_1}^{(k)}, {\mathbf{m}_2}^{(k)}, ..., {\mathbf{m}_{q^+}}^{(k)}\}$. We should find a permutation of \{$1, 2, ..., q^+$\}, denoted by 
\{$p_1, p_2, ..., p_{q^+}$\}, such that $F({\mathbf{y}_{p_i}}^{(k)})={\mathbf{m}_i}^{(k)}$ for $1\leq i\leq q^+$. It relies on the statistical distribution of 
${\mathbf{x}}$ and ${\mathbf{m}_i}^{(k)}~(1\leq i\leq q^+)$. It can be modeled as a \emph{minimum weight maximum matching (MWMM)} 
problem, which will be presented in near future.





%

\end{document}